\documentstyle[amssymb,preprint,aps]{revtex}

\newtheorem{theorem}{Theorem}
\newtheorem{acknowledgement}[theorem]{Acknowledgement}

\begin{document}
\title{Transport properties of fermionic systems}
\author{E. Prodan}
\address{University of Houston, Department of Physics, 4800 Calhoun Road\\
Houston, TX, 77204-5506}
\maketitle

\begin{abstract}
We extend the method discovered by A Y Alekseev et al \cite{F1} to the case
of fermions in external fields. A general formula for conductance $G$ is
proved. In the (1+1)-D case with symmetry at time reflection, it is shown
that: $G=e^{2}/h$ $+o\left( \alpha ^{2}\right) $, where $\alpha $ is the
strength of the external field. In (3+1)-D free case, it is checked that $%
G=ne^{2}/h$, where $n$ is the number of the filled energetic bands of the
transversal quantization.
\end{abstract}

\tightenlines

\section{Introduction}

\noindent It was proven relative recently that the equilibrium statistical
mechanics may be used at the description of some transport phenomena in
solids. Using this approach, A Y Alekseev et al \cite{F1} have found a
compact formula for conductance. We apply in this paper their method to
study the transport properties of the fermionic systems. The method have
been applied to the cases where the backscattering processes were not
present. In this paper we show how to incorporate in the theory the
backscattering processes. The presence of impurities can be modeled by
introducing an external field. To analyze fermionic systems in external
fields we make use of the implementation of the one particle picture. This
will enforce us to work only in the fermionic representation.

We calculate first the conductance of the free (1+1)-D fermionic system and
rediscover the results of \cite{F1}, obtained by using the bosonic
representation. Then we find a natural way to extend the formalism to the
case of the fermionic systems in external fields. A general formula for
conductance will be deduced. It is shown that the first correction to the
conductance is quadratic in the strength of the external field. This is
consistent with the fact that the conductance should be lower than $e^{2}/h$
for both cases: attractive and repelling external fields, because the
backscattering processes are present in both cases. However, because of the
complicated analytical expressions, we are not able to check that indeed the
conductance is strictly lower than $e^{2}/h$. We anticipate interesting
behavior of the conductance with the increase of the external field
strength: discontinuities when bound states appear or disappear.

The last section treats the (3+1)-D fermionic systems. It was argued in the
literature \cite{La} that the conductance of a pure quantum wire is
proportional with the number of filled energy bands of transversal
quantization. We check this for the free (3+1)-D fermionic system. For
(3+1)-D fermionic systems with external interactions only in the directions
perpendicular to the measured current, one can combine our results with
those of \cite{Si} to study the quantization of the conductance for Quantum
Hall effect.

\section{The general prescription}

\noindent We will briefly discuss first the method discovered by A Y
Alekseev et al in \cite{F1}. Given a quantum theory described by a
Hamiltonian ${\bf H}$, the first step is the identification of two physical
components of the quantum field, the left and right parts, and the
corresponding charges: $Q_{L,R}$. They must satisfy the following
commutation relations: 
\begin{equation}
\left[ {\bf H}\,,\,Q_{L,R}\right] =0\text{ and }\left[ Q_{L}\,,\,Q_{R}\right]
=0\text{.}
\end{equation}
Denoting the Noether conserved currents corresponding to the charges $%
Q_{L,R} $ by $j_{L,R}$ and the total current by $j$, the expectation value
of the current, in the equilibrium state corresponding to the chemical
potentials $\mu _{L,R}$ (denoted by $\left\langle \cdot \right\rangle _{\mu
} $), is: 
\begin{equation}
\left\langle {\bf j}\left( y\right) \right\rangle _{\mu }=-%
{\displaystyle{ie \over 2\hbar }}%
\left( \mu _{L}-\mu _{R}\right) \int dx\,\left\langle \left[ j_{a}^{0}\left(
x\right) ,{\bf a}\left( y\right) \right] \right\rangle _{\mu }\text{.}
\label{j}
\end{equation}
Here, ${\bf a}$ is the solution of: 
\begin{equation}
j^{0}=e\nabla {\bf a\,}\,\text{,\thinspace \thinspace \thinspace }{\bf j}%
=e\partial _{t}{\bf a}\text{,}
\end{equation}
and $j_{a}$ denotes the axial current, $j_{L}-j_{R}$. The electric potential
can be expressed as: $eV=\mu _{L}-\mu _{R}$.

In the following, we will refer to the above method as the general
prescription. If one works in the fermionic representation, the commutator $%
\left[ \rho _{a}\left( x\right) ,{\bf a}\left( y\right) \right] $ cannot be
calculated directly and we apply the following strategy. We calculate first
the expectation value: $f\left( x,y\right) =\left\langle \left[ \rho
_{a}\left( x\right) \,,\,\rho \left( y\right) \right] \right\rangle _{\mu }$%
. Then we search for a vector $\vec{F}\left( x,y\right) $ such that: $\vec{%
\bigtriangledown}_{y}\vec{F}\left( x,y\right) =f\left( x,y\right) $. In the
case of $\lim\limits_{\left| y\right| \rightarrow \infty }\vec{F}\left(
x,y\right) $ exists (and is well defined in the sense that it does not
depend on the direction), the expectation value of the commutator $\left[
\rho _{a}\left( x\right) ,{\bf a}\left( y\right) \right] $ is given by $\vec{%
F}\left( x,y\right) -\lim\limits_{\left| y\right| \rightarrow \infty }\vec{F}%
\left( x,y\right) $. This assures that $\rho _{a}\left( x\right) $ and ${\bf %
a}\left( y\right) $ decuple when $x$ and $y$ are separated by infinite large
distance, which must be the case if we assume that ${\bf a}$ is a local
observable. Along this paper, we will keep the notation $f$ and $\vec{F}$
for the above quantities.

\section{(1+1)-D fermionic systems}

\subsection{(1+1)-D free fermionic system}

\noindent Let us consider the free fermionic system described by the
Hamiltonian: 
\begin{equation}
{\bf H}_{0}=\sum_{k}%
{\displaystyle{k^{2} \over 2m}}%
\,\Psi _{k}^{\dagger }\,\Psi _{k}\text{,}
\end{equation}
for some finite length, $L$, and periodic boundary conditions. Following the
general prescription, the first step is to define the left and right
components of the field. We propose the zero time left and right fields and
the corresponding charge operators to be: 
\begin{equation}
\Psi _{L,R}\left( x\right) =\sum_{k\gtrless 0}%
{\displaystyle{e^{ikx} \over \sqrt{L}}}%
\Psi _{k}\text{ , }Q_{L,R}=\sum_{k\gtrless 0}\Psi _{k}^{\dagger }\,\Psi _{k}%
\text{.}
\end{equation}
It is easily to check that the following commutation relations held: 
\begin{equation}
\left[ {\bf H}_{0}\,,\,Q_{L,R}\right] =0,\text{ }\left[ Q_{L}\,,\,Q_{R}%
\right] =0\text{.}
\end{equation}
So, we are in the conditions required by the general prescription. The axial
charge density, $\rho _{a}$, is defined by: 
\begin{equation}
Q_{L}-Q_{R}=\int dx\,\left( \rho _{L}\left( x\right) -\rho _{R}\left(
x\right) \right) \equiv \int dx\,\rho _{a}\left( x\right) \text{.}
\end{equation}
The left and right charge densities are not uniquely defined. We choose the
following representation: 
\begin{equation}
\rho _{L,R}\left( x\right) =%
{\textstyle{1 \over 2}}%
\left[ \Psi ^{\dagger }\left( \Pi _{L,R}x\right) \Psi \left( x\right) +\Psi
^{\dagger }\left( x\right) \Psi \left( \Pi _{L,R}x\right) \right] 
\end{equation}
where $\Pi _{L,R}$ represent the projectors on the left states: $k>0$,
respectively on the right states: $k<0$. The notation is coming from:
\[
\Psi _{L,R}\left( x\right) =%
\mathop{\textstyle\sum}%
_{k\gtrless 0}%
{\displaystyle{e^{ikx} \over \sqrt{L}}}%
\Psi _{k}=%
\mathop{\textstyle\sum}%
_{k\gtrless 0}\int dy\,%
{\displaystyle{e^{ik\left( x-y\right) } \over L}}%
\Psi \left( y\right) 
\]
\begin{equation}
=\int dy\,\Pi _{L,R}\left( y,x\right) \Psi \left( y\right) =\Psi \left(
q\right) \text{,}
\end{equation}
where $q$ is given by: $q\left( y\right) =\left[ \Pi _{L,R}\delta \left(
x-\cdot \right) \right] \left( y\right) $ and we have denoted: $\Pi
_{L,R}\delta \left( x-\cdot \right) =\Pi _{L,R}x$. The zero temperature
expectation values corresponding to the chemical potentials $\mu _{L,R}=\mu
\pm 
{\displaystyle{eV \over 2}}%
$, can be calculated by employing the functional integral over the
Grassmanian variables with the measure corresponding to the covariance: 
\begin{equation}
C_{kk^{\prime }}=\delta _{kk^{\prime }}%
{\displaystyle{1 \over ik_{0}-\left[ k^{2}/2m-\mu _{L}\,\chi _{>}\left( k\right) -\mu _{R}\,\chi _{<}\left( k\right) \right] }}%
\text{,}
\end{equation}
where $\chi _{\lessgtr }$ are the characteristic functions of $\left(
-\infty ,0\right] $ and $\left[ 0,\infty \right) $ respectively. It follows
from Appendix A that the expectation value of the following commutator is
given by: 
\begin{equation}
\left\langle \left[ \rho _{a}\left( x\right) ,\rho \left( y\right) \right]
\right\rangle _{\mu }=i%
\mathop{\rm Im}%
\left[ \left\{ \Psi \left( \left( \Pi _{L}-\Pi _{R}\right) x\right)
\,,\,\Psi ^{\dagger }\left( y\right) \right\} \,\left\langle \Psi ^{\dagger
}\left( x\right) \Psi \left( y\right) \right\rangle _{\mu }\right] 
\label{com}
\end{equation}
and further: 
\begin{equation}
\left\langle \left[ \rho _{a}\left( x\right) ,\rho \left( y\right) \right]
\right\rangle _{\mu }=i%
\mathop{\rm Im}%
\left[ \left( \Pi _{L}-\Pi _{R}\right) \left( y,x\right) \,\left( \Pi
_{L}\,E_{\left[ 0,\mu _{L}\right] }+\Pi _{R}\,E_{\left[ 0,\mu _{R}\right]
}\right) \left( x,y\right) \right] \text{.}  \label{pro}
\end{equation}
We have denoted by $E_{\Omega }$ the spectral projection of $-%
{\displaystyle{\hbar ^{2} \over 2m}}%
\Delta _{P}$ ($\Delta _{P}$ the Laplace operator on $L$ with periodic
boundary conditions) for some $\Omega \in R$. In the infinite limit ($%
L\rightarrow \infty $), we have: 
\begin{equation}
f\left( x,y\right) =%
{\displaystyle{-i \over 2\pi ^{2}}}%
{\displaystyle{\sin k_{L}\left( x-y\right) +\sin k_{R}\left( x-y\right)  \over \left( x-y\right) ^{2}}}%
\text{.}
\end{equation}
Following our strategy, we find: 
\[
F\left( x,y\right) =%
{\displaystyle{-i \over 2\pi ^{2}}}%
\left\{ k_{L}\,%
\mathop{\rm Ci}%
\left[ k_{L}\left( x-y\right) \right] -%
{\displaystyle{\sin k_{L}\left( x-y\right)  \over x-y}}%
\right\} 
\]
\begin{equation}
+%
{\displaystyle{-i \over 2\pi ^{2}}}%
\left\{ k_{R}\,%
\mathop{\rm Ci}%
\left[ k_{R}\left( x-y\right) \right] -%
{\displaystyle{\sin k_{R}\left( x-y\right)  \over x-y}}%
\right\} \text{,}
\end{equation}
with the trivial limits $\lim\limits_{y\rightarrow \pm \infty }F\left(
x,y\right) =0$. From formula \ref{j}, the current density is: 
\begin{equation}
\left\langle j\left( y\right) \right\rangle _{\mu }=-%
{\displaystyle{i \over 2\hbar }}%
\left( \mu _{L}-\mu _{R}\right) \int dx\,F\left( x,y\right) =\left( \mu
_{L}-\mu _{R}\right) 
{\displaystyle{e \over h}}%
\end{equation}
which leads to $G=I/V=e^{2}/h$. The main goal of this subsection was to
check the validity of our strategy.

\subsection{(1+1)-D fermionic systems with external interactions}

\noindent We will consider in this subsection (1+1)-D fermionic systems in
external fields. We restrict here at the second quantization of the one
particle problems described by the Hamiltonians of the following type: $H=-%
{\displaystyle{\hbar ^{2} \over 2m}}%
\Delta +V$, considered again on a finite length, $L$, and with periodic
boundary conditions. We will consider first a compact supported potential, $%
suppV\subset \left[ a,b\right] $, with $-L/2<a<b<L/2$. Also, we consider the
case when the Hamiltonian $H$ doesn't have any bounded state in the limit $%
L\rightarrow \infty $. In the above conditions, the eigenvalues of the
Hamiltonian are again $%
{\displaystyle{k_{n}^{2} \over 2m}}%
$, with $k_{n}$ in a discrete set of numbers. Let $e_{k}$ be the eigenvector
corresponding to some $k$ with $e_{k}\left( x\right) \propto e^{-ikx}$ for $%
x<a$, $k>0$ and $e_{k}\left( x\right) \propto e^{ikx}$ for $x>b$, $k<0$ (the
``scattered'' states from left and right \cite{Lu}). The second quantized
Hamiltonian is: 
\begin{equation}
{\bf H}=\sum_{k}%
{\displaystyle{k^{2} \over 2m}}%
\,\Psi _{k}^{\dagger }\Psi _{k}\text{,}
\end{equation}
where $\Psi _{k}$ has now the following meaning: 
\begin{equation}
\Psi _{k}=\int dx\,e_{k}\left( x\right) \Psi \left( x\right) \text{.}
\end{equation}
The left and right components of the field and the corresponding charge
operators can be defined analogous: 
\begin{equation}
\Psi _{L,R}\left( x\right) =\sum_{k\gtrless 0}e_{k}^{\ast }\left( x\right)
\Psi _{k}=\Psi \left( \Pi _{L,R}x\right) \text{, }Q_{L,R}=\sum_{k\gtrless
0}\Psi _{k}^{\dagger }\Psi _{k}\text{,}
\end{equation}
where $\Pi _{L,R}$ are again the projectors on the states with $k>0$ and $k<0
$ respectively. It is straightforward that the charge operators commute with
the Hamiltonian. So we are again in the conditions of the general
prescription. The current density will be given by the same expression \ref
{j}. Further, because $\left\{ \Psi _{k}\right\} $ satisfy the same
anti-commutation relations we can conclude that: 
\begin{equation}
\left\langle \left[ \rho _{a}\left( x\right) ,\rho \left( y\right) \right]
\right\rangle _{\mu }=i%
\mathop{\rm Im}%
\left[ \left\{ \Psi \left( \left( \Pi _{L}-\Pi _{R}\right) x\right)
\,,\,\Psi ^{\dagger }\left( y\right) \right\} \,\left\langle \Psi ^{\dagger
}\left( x\right) \Psi \left( y\right) \right\rangle _{\mu }\right] \text{.}
\end{equation}
According to Appendix B, the above expression can be written completely
analogous as: 
\begin{equation}
\left\langle \left[ \rho _{a}\left( x\right) ,\rho \left( y\right) \right]
\right\rangle =i%
\mathop{\rm Im}%
\left[ \left( \Pi _{L}-\Pi _{R}\right) \left( y,x\right) \,\left( \Pi
_{L}\,E_{\left[ 0,\mu _{L}\right] }+\Pi _{R}\,E_{\left[ 0,\mu _{R}\right]
}\right) \left( x,y\right) \right] \text{,}
\end{equation}
where $E_{\Omega }$ is the spectral projector of $H$ corresponding to some $%
\Omega \in R$. In the infinite limit ($L\rightarrow \infty $), we can use
the results of the last section if a supplementary condition is fulfilled,
which is: the Moller operators $W_{\pm }=s-\lim\limits_{T\rightarrow \pm
\infty }e^{iTH}e^{-iTH_{0}}$ exist and are complete. In this case one can
use the interwining properties of Moller operators, $E_{\Omega
}=W_{+}\,E_{\Omega }^{\left( 0\right) }\,W_{+}^{\dagger }$ and the obvious
relation: $\Pi _{L,R}=W_{+}\,\Pi _{L,R}^{\left( 0\right) }\,W_{+}^{\dagger }$%
. In more complicate situations, the last relation can be used as the
definition of the left and right states. The index $\left( 0\right) $ is
referring to the free case. Using the explicit expressions of the projectors
for the free case, one has: 
\begin{equation}
\left\langle \left[ \rho _{a}\left( x\right) ,\rho \left( y\right) \right]
\right\rangle _{\mu }=i%
\mathop{\rm Im}%
\left\{ \left[ W_{+}\,\hat{A}\,W_{+}^{\dagger }\right] \left( y,x\right) \,%
\left[ W_{+}\,\hat{B}\,W_{+}^{\dagger }\right] \left( x,y\right) \right\} 
\text{,}
\end{equation}
where $\hat{A}$ and $\hat{B}$ are the operators corresponding to the
kernels: 
\begin{equation}
A\left( x,y\right) =%
{\displaystyle{1 \over i\pi \left( x-y\right) }}%
\,\text{,\thinspace\ }B\left( x,y\right) =%
{\displaystyle{\sin \left[ %
{\textstyle{k_{L}+k_{R} \over 2}}\left( x-y\right) \right]  \over \pi \left( x-y\right) }}%
e^{-i\frac{k_{L}-k_{R}}{2}\left( x-y\right) }\text{.}
\end{equation}
One can go one step further by employing the following argument. If one
changes the sign of the applied electric voltage, then the new current must
have the same intensity but opposite direction. This is due to symmetry at
time reflection. It implies that $\left\langle j\left( y\right)
\right\rangle _{\mu }$ must be anti-symmetric at the interchange: $%
k_{L}\rightleftharpoons k_{R}$. Because in the formula \ref{j} there is
already the term $\left( \mu _{L}-\mu _{R}\right) $ in front of the
integral, it follows that we should retain in $\left\langle \left[ \rho
_{a}\left( x\right) ,\rho \left( y\right) \right] \right\rangle _{\mu }$
only the symmetric term relativ to the interchange: $k_{L}\rightleftharpoons
k_{R}$. From the expression of $B\left( x,y\right) $ it follows: 
\begin{equation}
\left\langle \left[ \rho _{a}\left( x\right) ,\rho \left( y\right) \right]
\right\rangle _{\mu }=i%
\mathop{\rm Im}%
\left\{ \left[ W_{+}\,\hat{A}\,W_{+}^{\dagger }\right] \left( y,x\right) \,%
\left[ W_{+}\,%
\mathop{\rm Re}%
\hat{B}\,W_{+}^{\dagger }\right] \left( x,y\right) \right\} \text{.}
\label{fo}
\end{equation}
We consider that the formula (\ref{fo}) can be used to calculate the
conductance for more general cases (for example when bounded states are
present). Indeed, the transport of electric charges can be done only by
scattering states. One can see that the formula (\ref{fo}) takes into
account only the scattering states so it can be used directly without going
back and redefine the currents (which should incorporate only the scattering
states).

\subsection{The dependence on the strength of the external field}

\noindent Let us introduce a scaling factor, $\alpha $, in front of the
potential: $V\rightarrow \alpha V$. We study in this subsection the
dependence of the conductance on the parameter $\alpha $. The principal
result is: $\left. dG/d\alpha \right| _{\alpha =0}=0$ which leads to the
following expansion of $G$: 
\begin{equation}
G=%
{\displaystyle{e^{2} \over h}}%
+G^{\left( 2\right) }\alpha ^{2}+o\left( \alpha ^{3}\right) .
\end{equation}
One could anticipate this by considering both cases, $\pm V$. If the first
perturbative correction were non-zero, then, for one of these two cases, the
conductance were greater than that of the free case which is wrong because
in both cases the backscattering processes are present. This argument also
implies that $G^{\left( 2\right) }$ should be a negative number.

The stationary representation of the Moller operator is \cite{Ku}: 
\[
W_{+}=I-%
{\displaystyle{\alpha  \over 2\pi i}}%
\lim\limits_{\varepsilon \searrow 0}\int\limits_{-\infty }^{\infty
}R_{0}\left( \lambda -i\varepsilon \right) \,V\,R_{0}\left( \lambda
+i\varepsilon \right) \,d\lambda 
\]
\[
+%
{\displaystyle{\alpha ^{2} \over 2\pi i}}%
\lim\limits_{\varepsilon \searrow 0}\int\limits_{0}^{\infty }R_{0}\left(
\lambda -i\varepsilon \right) \,V\,R_{\alpha }\left( \lambda -i\varepsilon
\right) \,V\,\left\{ R_{0}\left( \lambda +i\varepsilon \right) -R_{0}\left(
\lambda -i\varepsilon \right) \right\} d\lambda 
\]
\begin{equation}
\equiv I+\alpha W_{1}+\alpha ^{2}W_{2}\text{,}
\end{equation}
where $R_{\alpha }\left( z\right) $ represents the resolvent of $-\Delta
+\alpha V$. Then one can see that: 
\begin{equation}
\left. 
{\displaystyle{d \over d\alpha }}%
\left[ \rho _{a}\left( x\right) ,\rho \left( y\right) \right] \right|
_{\alpha =0}=i%
\mathop{\rm Im}%
\left\{ \left[ W_{1}\,,\,\hat{A}\right] \left( y,x\right) \,B\left(
x,y\right) +A\left( y,x\right) \left[ W_{1}\,,\,\hat{B}\right] \left(
x,y\right) \right\} \text{.}
\end{equation}
With the observation of the last subsection: 
\begin{equation}
\left. 
{\displaystyle{d \over d\alpha }}%
\left[ \rho _{a}\left( x\right) ,\rho \left( y\right) \right] \right|
_{\alpha =0}=\left[ 
\mathop{\rm Re}%
W_{1}\,,\,\hat{A}\right] \left( y,x\right) \,%
\mathop{\rm Re}%
\left[ B\left( x,y\right) \right] +A\left( y,x\right) \left[ 
\mathop{\rm Re}%
W_{1}\,,\,%
\mathop{\rm Re}%
\hat{B}\right] \left( x,y\right)   \label{car}
\end{equation}
In our further calculations, we will deduce for $W_{1}$ a formula similar to
that of \cite{Ya}. The calculations are lengthy and are presented in the
Appendix C, where it is proved that $\left. 
{\displaystyle{d \over d\alpha }}%
\left[ \rho _{a}\left( x\right) ,\rho \left( y\right) \right] \right|
_{\alpha =0}=0$.

The coefficients of the upper powers of $\alpha $ are to complicated (at
least for the author) to be handled analytically. However, a numerical
approach will not be a trivial thing and it can make alone the subject of a
new study.

\subsection{Comments}

\noindent We have pointed out before, that the formula \ref{fo} can be used
in more general context. The Moller operators project out the bounded
statets (and zero energy resonances). As a result, only the scattering
states contribute to the current. One can see that there is a major
difference between the cases with and without bounded states. We expect
discontinuities of the conductance (versus strength of the external field)
when bounded states are forming. Thus, the investigation of the transport
properties may be a complementary way to the method of scatterings in
detection and measuring of the spectrum of energy for quantum systems.

\section{(3+1)-D fermionic systems}

The primary goal of this section is to prove that the conductance of the
free (3+1)-D fermionic system is proportional with the number of filled
energy bands of the transversal quantization. The left and right fields and
corresponding charge operators can be chosen as: 
\begin{equation}
\Psi _{L,R}\left( x\right) =%
{\displaystyle{1 \over \sqrt{v}}}%
\sum_{k_{1}\gtrless 0}e^{i\vec{k}\,\vec{x}}\Psi _{\vec{k}}\,,\,\,Q_{L,R}=%
\sum_{k_{1}\gtrless 0}\Psi _{\vec{k}}^{\dagger }\,\Psi _{\vec{k}}\text{,}
\end{equation}
where $x_{1}$ denotes the direction of the measured current and $v$ is the
volume. One can follow the same steps as in (1+1)-D case to prove the
validity of the formula \ref{com}. According to Appendix D, in the infinite
limite: 
\begin{equation}
f\left( x,y\right) =%
{\displaystyle{i\delta \left( x_{\bot }-y_{\bot }\right)  \over \pi }}%
\left\{ k_{L}^{4}\,u\left( k_{L}\left( x_{1}-y_{1}\right) \right)
+k_{R}^{4}\,u\left( k_{R}\left( x_{1}-y_{1}\right) \right) \right\} \text{,}
\end{equation}
where $u\left( x\right) =%
{\displaystyle{\cos x \over x^{3}}}%
-%
{\displaystyle{\sin x \over x^{4}}}%
$. Because the current flows only in $x_{1}$ direction, we can choose the
vector ${\bf a}$ along $x_{1}$ and we can solve easily the equation $\vec{%
\bigtriangledown}_{y}\vec{F}=f\left( x,y\right) $: 
\begin{equation}
F_{1}\left( x,y\right) =\left\langle \left[ \rho _{a}\left( x\right)
,a_{1}\left( y\right) \right] \right\rangle _{\mu }=%
{\displaystyle{i\delta \left( x_{\bot }-y_{\bot }\right)  \over \pi }}%
\left\{ k_{L}^{3}\,U\left( k_{L}\left( x_{1}-y_{1}\right) \right)
+k_{R}^{3}\,U\left( k_{R}\left( x_{1}-y_{1}\right) \right) \right\} \text{,}
\end{equation}
with $U\left( x\right) =-%
{\displaystyle{\cos x \over 3x^{2}}}%
-%
{\displaystyle{%
\mathop{\rm Ci}x \over 3}}%
+%
{\displaystyle{\sin x \over 3x^{3}}}%
+%
{\displaystyle{\sin x \over 3x}}%
$. Using that $\int_{-\infty }^{\infty }U\left( x\right) dx=\pi /2$, it
follows: 
\begin{equation}
\left\langle j_{1}\left( y\right) \right\rangle _{\mu }=\left\langle \left[
Q_{L}-Q_{R},a_{1}\left( y\right) \right] \right\rangle _{\mu }=\int dx_{\bot
}\int dx_{1}\left\langle \left[ \rho _{a}\left( x\right) ,a\left( y\right) %
\right] \right\rangle =%
{\displaystyle{i \over 2}}%
\left( k_{L}^{2}+k_{R}^{2}\right) =ik_{F}^{2}
\end{equation}
which leads to: 
\begin{equation}
I=\int_{\Delta S}dy_{\bot }\left\langle j_{1}\left( y\right) \right\rangle =-%
{\displaystyle{ie \over 2\hbar }}%
\left( \mu _{L}-\mu _{R}\right) ik_{F}^{2}\Delta S=%
{\displaystyle{e^{2} \over h}}%
\left( \pi k_{F}^{2}\Delta S\right) \cdot V\text{.}
\end{equation}
Thus: $G=I/V=\left( \pi k_{F}^{2}\Delta S\right) \,%
{\displaystyle{e^{2} \over h}}%
$. Of course, the pre-factor $\pi k_{F}^{2}\Delta S$ represents the number
of the transversal conducting channels. The result may be generalized to the
case when transversal interactions are included. In \cite{Si} it was
calculated the change in the number of energetic bands below the Fermi
energy for 2-D quantum systems, when infinitesimally thin magnetic flux
tubes are introduced. We think that the two results can be combined to study
the quantization of the conductance for Quantum Hall effect.\smallskip 

\begin{acknowledgement}
The author gratefully acknowledges support, under the direction of J.
Miller, by the State of Texas through the Texas Center for Superconductivity
and the Texas Higher Education Coordinating Board Advanced Technology
Program, and by the Robert A. Welch Foundation.
\end{acknowledgement}

\section{Appendix A}

\noindent In this section we will calculate explicitly the commutator for
the free fermions: 
\[
\left[ \rho _{a}\left( x\right) ,\rho \left( y\right) \right] =%
{\displaystyle{1 \over 2}}%
\left[ \Psi ^{\dagger }\left( \left( \Pi _{L}-\Pi _{R}\right) x\right)
\,\Psi \left( x\right) \,,\,\Psi ^{\dagger }\left( y\right) \Psi \left(
y\right) \right] 
\]
\begin{equation}
+%
{\displaystyle{1 \over 2}}%
\left[ \Psi ^{\dagger }\left( x\right) \Psi \left( \left( \Pi _{L}-\Pi
_{R}\right) x\right) \,,\,\Psi ^{\dagger }\left( y\right) \Psi \left(
y\right) \right] \text{.}
\end{equation}
We can use here the general formula: 
\begin{equation}
\left[ ab\,,\,cd\right] =a\left[ \left\{ b\,,\,c\right\} d-c\left\{
b\,,\,d\right\} \right] +\left[ \left\{ a\,,\,c\right\} d-c\left\{
a\,,\,d\right\} \right] b
\end{equation}
which leads to: 
\[
\left[ \Psi ^{\dagger }\left( \left( \Pi _{L}-\Pi _{R}\right) x\right)
\,\Psi \left( x\right) \,,\,\Psi ^{\dagger }\left( y\right) \Psi \left(
y\right) \right] =\delta \left( x-y\right) \Psi ^{\dagger }\left( \left( \Pi
_{L}-\Pi _{R}\right) x\right) \Psi \left( y\right) 
\]
\begin{equation}
-\left\{ \Psi ^{\dagger }\left( \left( \Pi _{L}-\Pi _{R}\right) x\right)
\,,\,\Psi \left( y\right) \right\} \,\Psi ^{\dagger }\left( y\right) \Psi
\left( x\right)
\end{equation}
and 
\[
\left[ \Psi ^{\dagger }\left( x\right) \Psi \left( \left( \Pi _{L}-\Pi
_{R}\right) x\right) \,,\,\Psi ^{\dagger }\left( y\right) \Psi \left(
y\right) \right] =\left\{ \Psi \left( \left( \Pi _{L}-\Pi _{R}\right)
x\right) \,,\,\Psi ^{\dagger }\left( y\right) \right\} \,\Psi ^{\dagger
}\left( x\right) \Psi \left( y\right) 
\]
\begin{equation}
-\delta \left( x-y\right) \Psi ^{\dagger }\left( y\right) \Psi \left( \left(
\Pi _{L}-\Pi _{R}\right) x\right) \text{.}
\end{equation}
In consequence, the commutator can be expressed as: 
\[
\left[ \rho _{a}\left( x\right) ,\rho \left( y\right) \right] =%
{\displaystyle{1 \over 2}}%
\delta \left( x-y\right) \left[ \Psi ^{\dagger }\left( \left( \Pi _{L}-\Pi
_{R}\right) x\right) \Psi \left( y\right) -\Psi ^{\dagger }\left( y\right)
\Psi \left( \left( \Pi _{L}-\Pi _{R}\right) x\right) \right] 
\]
\[
+%
{\displaystyle{1 \over 2}}%
\left\{ \Psi \left( \left( \Pi _{L}-\Pi _{R}\right) x\right) \,,\,\Psi
^{\dagger }\left( y\right) \right\} \,\Psi ^{\dagger }\left( x\right) \Psi
\left( y\right) 
\]
\begin{equation}
-%
{\displaystyle{1 \over 2}}%
\left\{ \Psi ^{\dagger }\left( \left( \Pi _{L}-\Pi _{R}\right) x\right)
\,,\,\Psi \left( y\right) \right\} \,\Psi ^{\dagger }\left( y\right) \Psi
\left( x\right) \text{.}
\end{equation}
By a simple operation, one can put the first term in the form: 
\begin{equation}
\delta \left( x-y\right) \left[ \Psi _{L}^{\dagger }\left( x\right) \Psi
_{R}\left( x\right) -\Psi _{R}^{\dagger }\left( x\right) \Psi _{L}\left(
x\right) \right]
\end{equation}
and one can see that its expectation value is zero. If now one considers the
expectation value of the next two terms, the result can be written in the
compact form: 
\begin{equation}
\left\langle \left[ \rho _{a}\left( x\right) ,\rho \left( y\right) \right]
\right\rangle _{\mu }=i%
\mathop{\rm Im}%
\left\{ \Psi \left( \left( \Pi _{L}-\Pi _{R}\right) x\right) \,,\,\Psi
^{\dagger }\left( y\right) \right\} \left\langle \Psi ^{\dagger }\left(
x\right) \Psi \left( y\right) \right\rangle _{\mu }
\end{equation}
by taking into account that the anti-commutator is just a number. Moreover,
by using the field anti-commutation relations: 
\begin{equation}
\left\{ \Psi \left( \left( \Pi _{L}-\Pi _{R}\right) x\right) \,,\,\Psi
^{\dagger }\left( y\right) \right\} =%
{\displaystyle{1 \over L}}%
\sum_{k>0}e^{ik\left( x-y\right) }-%
{\displaystyle{1 \over L}}%
\sum_{k<0}e^{ik\left( x-y\right) }=\left( \Pi _{L}-\Pi _{R}\right) \left(
y,x\right)
\end{equation}
which in the continuous limit becomes: 
\begin{equation}
{\displaystyle{1 \over 2\pi }}%
\int_{-\infty }^{\infty }dk\,\left( \chi _{>}\left( k\right) -\chi
_{<}\left( k\right) \right) e^{ik\left( x-y\right) }=%
{\displaystyle{1 \over 2\pi ^{2}i}}%
\int_{-\infty }^{\infty }dk\int_{-\infty }^{\infty }d\eta 
{\displaystyle{e^{ik\eta } \over \eta }}%
e^{ik\left( x-y\right) }=%
{\displaystyle{-1 \over i\pi \left( x-y\right) }}%
\text{.}
\end{equation}
The propagator can be calculated by using the functional integral: 
\begin{equation}
\left\langle \Psi ^{\dagger }\left( x\right) \Psi \left( y\right)
\right\rangle _{\mu }=\int \Psi ^{\dagger }\left( x\right) \Psi \left(
y\right) d\mu _{C}
\end{equation}
with Grassmann Gaussian measure corresponding to covariance: 
\begin{equation}
C_{k,k^{\prime }}=\delta _{k,k^{\prime }}%
{\displaystyle{1 \over ik_{0}-\left( k^{2}/2m-\mu _{L}\,\chi _{>}\left( k\right) -\mu _{R}\,\chi _{<}\left( k\right) \right) }}%
\text{.}
\end{equation}
\[
\left\langle \Psi ^{\dagger }\left( x\right) \Psi \left( y\right)
\right\rangle _{\mu }=\sum_{k}\Theta \left( k^{2}/2m-\mu _{L}\,\chi
_{>}\left( k\right) -\mu _{R}\,\chi _{<}\left( k\right) \right) 
{\displaystyle{e^{-ik\left( x-y\right) } \over L}}%
\]
\begin{equation}
=\left( \Pi _{L}E_{\left[ 0,\mu _{L}\right] }+\Pi _{R}E_{\left[ 0,\mu _{R}%
\right] }\right) \left( x,y\right) \text{.}
\end{equation}
In the infinite limit, the propagator becomes: 
\begin{equation}
\left\langle \Psi ^{\dagger }\left( x\right) \Psi \left( y\right)
\right\rangle _{\mu }=\int\limits_{-k_{R}}^{k_{L}}%
{\displaystyle{dk \over 2\pi }}%
\,e^{-ik\left( x-y\right) }=%
{\displaystyle{\sin k_{F}\left( x-y\right)  \over \pi \left( x-y\right) }}%
e^{-i\frac{k_{L}-k_{R}}{2}\left( x-y\right) }\text{.}
\end{equation}

\section{Appendix B}

\noindent We prove here that the formula \ref{pro} is still valid for the
interacting case. Because the fields $\Psi _{k}$ satisfy the usual
commutation relations, $\left\{ \Psi _{k^{\prime }}\,,\,\Psi _{k}^{\dagger
}\right\} =\delta _{k,k^{\prime }}$, the formula \ref{com} results by
following the same steps. Now 
\[
\left\{ \Psi \left( \left( \Pi _{L}-\Pi _{R}\right) x\right) \,,\,\Psi
^{\dagger }\left( y\right) \right\} =\sum_{k,\,k^{\prime }>0}e_{k}\left(
y\right) e_{k^{\prime }}^{\ast }\left( x\right) \left\{ \Psi _{k^{\prime
}}\,,\,\Psi _{k}^{\dagger }\right\} -\sum_{k,\,k^{\prime }<0}e_{k}\left(
y\right) e_{k^{\prime }}^{\ast }\left( x\right) \left\{ \Psi _{k^{\prime
}}\,,\,\Psi _{k}^{\dagger }\right\} 
\]
\begin{equation}
=\sum_{\,k>0}e_{k}\left( y\right) e_{k}^{\ast }\left( x\right)
-\sum_{k<0}e_{k}\left( y\right) e_{k}^{\ast }\left( x\right) =\left( \Pi
_{L}-\Pi _{R}\right) \left( y,x\right) \text{.}
\end{equation}
For the expectation value $\left\langle \Psi ^{\dagger }\left( x\right) \Psi
\left( y\right) \right\rangle $ can be calculated in the same manner as in
Appendix A. The covariance of the Grassmann Gaussian measure for the
interacting case has the same matrix elements but relative to the
orthonormal basis $\left\{ e_{k}\right\} _{k}$. The functional integral
leads to: 
\[
\left\langle \Psi ^{\dagger }\left( x\right) \Psi \left( y\right)
\right\rangle _{\mu }=\sum_{k}\Theta \left( k^{2}/2m-\mu _{L}\,\chi
_{>}\left( k\right) -\mu _{R}\,\chi _{<}\left( k\right) \right) e_{k}\left(
x\right) e_{k}^{\ast }\left( y\right) 
\]
\begin{equation}
=\left( \Pi _{L}E_{\left[ 0,\mu _{L}\right] }+\Pi _{R}E_{\left[ 0,\mu _{R}%
\right] }\right) \left( x,y\right) \text{.}
\end{equation}
We can conclude that the expression \ref{pro} is still valid for this case.

\section{Appendix C}

We show here that $\left. 
{\displaystyle{dG \over d\alpha }}%
\right| _{\alpha =0}=0$. 
\[
W_{1}\left( x,y\right) =%
{\displaystyle{i \over \left( 2\pi \right) ^{3}}}%
\int_{-\infty }^{\infty }d\lambda \int_{-\infty }^{\infty }dz\int_{-\infty
}^{\infty }dk%
{\displaystyle{e^{-ik\left( x-z\right) } \over k^{2}-\left( \lambda -i\varepsilon \right) }}%
V\left( z\right) \int_{-\infty }^{\infty }dk^{\prime }%
{\displaystyle{e^{-ik^{\prime }\left( z-y\right) } \over k^{\prime 2}-\left( \lambda +i\varepsilon \right) }}%
\]
\begin{equation}
=-\left( 2\pi \right) ^{-3/2}\int_{-\infty }^{\infty }dk\int_{-\infty
}^{\infty }dk^{\prime }\,%
{\displaystyle{e^{-ikx}\tilde{V}\left( k^{\prime }-k\right) e^{ik^{\prime }y} \over k^{\prime 2}-k^{2}-2i\varepsilon }}%
\text{,}
\end{equation}
where \symbol{126} indicates the Fourier transform. We make the following
change of variables:

\begin{equation}
\left\{ 
\begin{array}{c}
\xi =k^{\prime }-k \\ 
\eta =k^{\prime }+k
\end{array}
\right.
\end{equation}
with the Jacobian: $%
{\displaystyle{\partial \left( k,k^{\prime }\right)  \over \partial \left( \xi ,\eta \right) }}%
=1/2$. Then: 
\[
W_{1}\left( x,y\right) =-%
{\displaystyle{\left( 2\pi \right) ^{-3/2} \over 2}}%
\int_{-\infty }^{\infty }d\xi \int_{-\infty }^{\infty }d\eta \,%
{\displaystyle{e^{i\xi \left( x+y\right) /2}\tilde{V}\left( \xi \right) e^{-i\eta \left( x-y\right) /2} \over \xi \eta -2i\varepsilon }}%
\]
\[
=-%
{\displaystyle{\left( 2\pi \right) ^{-3/2} \over 2}}%
\int_{-\infty }^{\infty }d\xi 
{\displaystyle{e^{i\xi \left( x+y\right) /2}\tilde{V}\left( \xi \right)  \over \xi }}%
\int_{-\infty }^{\infty }d\eta \,%
{\displaystyle{e^{-i\eta \left( x-y\right) /2} \over \eta -2i\varepsilon /\xi }}%
\]
\begin{equation}
=-%
{\displaystyle{i\left( 2\pi \right) ^{-1/2} \over 2}}%
\int_{-\infty }^{\infty }d\xi \,\left[ \chi _{>}\left( x-y\right) \chi
_{>}\left( \xi \right) -\chi _{<}\left( x-y\right) \chi _{<}\left( \xi
\right) \right] 
{\displaystyle{e^{i\xi \left( x+y\right) /2}\tilde{V}\left( \xi \right)  \over \xi }}%
\text{,}
\end{equation}
which is the expression of $W_{1}$ in \cite{Ya}. Further: 
\[
\mathop{\rm Re}%
W_{1}\left( x,y\right) =-%
{\displaystyle{i\left( 2\pi \right) ^{-1/2} \over 2}}%
\int_{-\infty }^{\infty }d\xi \,\left[ \chi _{>}\left( x-y\right) \chi
_{>}\left( \xi \right) -\chi _{<}\left( x-y\right) \chi _{<}\left( \xi
\right) \right] \times 
\]
\[
\times 
{\displaystyle{e^{i\xi \left( x+y\right) /2}\tilde{V}\left( \xi \right) -e^{-i\xi \left( x+y\right) /2}\tilde{V}\left( -\xi \right)  \over 2\xi }}%
\]
\[
=-%
{\displaystyle{i\left( 2\pi \right) ^{-1/2} \over 2}}%
\left[ \chi _{>}\left( x-y\right) -\chi _{<}\left( x-y\right) \right]
\int_{-\infty }^{\infty }d\xi \,\chi _{>}\left( \xi \right) 
{\displaystyle{e^{i\xi \left( x+y\right) /2}\tilde{V}\left( \xi \right) -e^{-i\xi \left( x+y\right) /2}\tilde{V}\left( -\xi \right)  \over 2\xi }}%
\]
\begin{equation}
=%
{\displaystyle{\left( 2\pi \right) ^{-1/2} \over 2}}%
\left[ \chi _{>}\left( x-y\right) -\chi _{<}\left( x-y\right) \right]
\int_{-\infty }^{\infty }d\xi \,%
{\displaystyle{e^{i\xi \left( x+y\right) /2}\tilde{V}\left( \xi \right)  \over 2i\xi }}%
\text{.}
\end{equation}
We write generically: 
\begin{equation}
\mathop{\rm Re}%
W_{1}\left( x,y\right) =\left[ \chi _{>}\left( x-y\right) -\chi _{<}\left(
x-y\right) \right] Q\left( x+y\right)
\end{equation}
The general expression for the commutators $\left[ 
\mathop{\rm Re}%
W_{1},\hat{C}\right] $, where $C\left( x,y\right) =C\left( x-y\right) $,
will be: 
\[
\left[ 
\mathop{\rm Re}%
W_{1},\hat{C}\right] \left( x,y\right) =\int_{-\infty }^{\infty }dz\,\left[
\chi _{>}\left( x-z\right) -\chi _{<}\left( x-z\right) \right] Q\left(
x+z\right) C\left( z-y\right) 
\]
\[
-\int_{-\infty }^{\infty }dz\,C\left( x-z\right) \left[ \chi _{>}\left(
z-y\right) -\chi _{<}\left( z-y\right) \right] Q\left( z+y\right) 
\]
\begin{equation}
=\int_{-\infty }^{\infty }dz\,\left[ \chi _{>}\left( x-y-z\right) -\chi
_{<}\left( x-y-z\right) \right] \left( Q\left( x+y+z\right) -Q\left(
x+y-z\right) \right) C\left( z\right)
\end{equation}
after a change of variable: $z\rightarrow z-y$ in the first integral and: $%
z\rightarrow -z+x$ in the second integral. Concretely: 
\begin{equation}
\left[ 
\mathop{\rm Re}%
W_{1},\hat{C}\right] \left( x,y\right) =%
{\textstyle{\left( 2\pi \right) ^{-1/2} \over 2}}%
\int_{-\infty }^{\infty }d\xi \,%
{\displaystyle{e^{i\xi \left( x+y\right) /2}\tilde{V}\left( \xi \right)  \over \xi }}%
\left( \int_{-\infty }^{x-y}-\int_{x-y}^{\infty }\right) dz\,\sin \left( 
{\displaystyle{\xi z \over 2}}%
\right) C\left( z\right) \text{.}
\end{equation}
It follows: 
\[
\left[ 
\mathop{\rm Re}%
W_{1},\hat{A}\right] \left( x,y\right) =%
{\textstyle{\left( 2\pi \right) ^{-1/2} \over 2}}%
\int_{-\infty }^{\infty }d\xi \,%
{\displaystyle{e^{i\xi \left( x+y\right) /2}\tilde{V}\left( \xi \right)  \over \xi }}%
\left( \int_{-\infty }^{x-y}-\int_{x-y}^{\infty }\right) dz\,%
{\displaystyle{\sin \left( %
{\displaystyle{\xi z \over 2}}\right)  \over i\pi z}}%
\]
\[
=%
{\textstyle{1 \over i\left( 2\pi \right) ^{3/2}}}%
\int_{-\infty }^{\infty }d\xi \,%
{\displaystyle{e^{i\xi \left( x+y\right) /2}\tilde{V}\left( \xi \right)  \over \xi }}%
\left[ 
\mathop{\rm Si}%
\left( x-y\right) -%
\mathop{\rm Si}%
\left( -\infty \right) -%
\mathop{\rm Si}%
\left( \infty \right) +%
\mathop{\rm Si}%
\left( x-y\right) \right] 
\]
\begin{equation}
=%
{\textstyle{2 \over i\left( 2\pi \right) ^{3/2}}}%
\int_{-\infty }^{\infty }d\xi \,%
{\displaystyle{e^{i\xi \left( x+y\right) /2}\tilde{V}\left( \xi \right)  \over \xi }}%
\mathop{\rm Si}%
\left( 
{\displaystyle{\xi  \over 2}}%
\left( x-y\right) \right)
\end{equation}
and

$\qquad \left[ 
\mathop{\rm Re}%
W_{1},%
\mathop{\rm Re}%
\hat{B}\right] \left( x,y\right) $%
\[
=%
{\textstyle{\left( 2\pi \right) ^{-1/2} \over 2}}%
\int_{-\infty }^{\infty }d\xi \,%
{\displaystyle{e^{i\xi \left( x+y\right) /2}\tilde{V}\left( \xi \right)  \over \xi }}%
\left( \int_{-\infty }^{x-y}-\int_{x-y}^{\infty }\right) dz\,\sin \left( 
{\displaystyle{\xi z \over 2}}%
\right) 
{\displaystyle{\sin \left( k_{L}z\right) +\sin \left( k_{R}z\right)  \over 2\pi z}}%
\]
\[
=%
{\textstyle{1 \over 2\left( 2\pi \right) ^{3/2}}}%
\int_{-\infty }^{\infty }d\xi \,%
{\displaystyle{e^{i\xi \left( x+y\right) /2}\tilde{V}\left( \xi \right)  \over \xi }}%
\left( \int_{-\infty }^{x-y}-\int_{x-y}^{\infty }\right) dz\,%
{\displaystyle{\cos \left[ \left( k_{L}-\xi /2\right) z\right] -\cos \left[ \left( k_{L}+\xi /2\right) z\right]  \over 2z}}%
\]
\[
+%
{\textstyle{1 \over 2\left( 2\pi \right) ^{3/2}}}%
\int_{-\infty }^{\infty }d\xi \,%
{\displaystyle{e^{i\xi \left( x+y\right) /2}\tilde{V}\left( \xi \right)  \over \xi }}%
\left( \int_{-\infty }^{x-y}-\int_{x-y}^{\infty }\right) dz\,%
{\displaystyle{\cos \left[ \left( k_{R}-\xi /2\right) z\right] -\cos \left[ \left( k_{R}+\xi /2\right) z\right]  \over 2z}}%
\]
\[
=%
{\textstyle{1 \over 2\left( 2\pi \right) ^{3/2}}}%
\int_{-\infty }^{\infty }d\xi \,%
{\displaystyle{e^{i\xi \left( x+y\right) /2}\tilde{V}\left( \xi \right)  \over \xi }}%
\left\{ 
\mathop{\rm Ci}%
\left[ \left( k_{L}-\xi /2\right) \left( x-y\right) \right] -%
\mathop{\rm Ci}%
\left[ \left( k_{L}+\xi /2\right) \left( x-y\right) \right] \right. 
\]
\begin{equation}
\left. +%
\mathop{\rm Ci}%
\left[ \left( k_{R}-\xi /2\right) \left( x-y\right) \right] -%
\mathop{\rm Ci}%
\left[ \left( k_{R}+\xi /2\right) \left( x-y\right) \right] \right\} \text{.}
\end{equation}
Now we start to apply the strategy of section 2. We need to calculate: 
\begin{equation}
\int_{-\infty }^{\infty }dx\int dy\left\{ \left[ W_{1}\,,\,\hat{A}\right]
\left( y,x\right) \,%
\mathop{\rm Re}%
\left[ B\left( x,y\right) \right] +A\left( y,x\right) 
\mathop{\rm Re}%
\left[ W_{1}\,,\,\hat{B}\right] \left( x,y\right) \right\} \text{.}
\label{fa}
\end{equation}
Changing the order of integration, one can see that the situation is similar
to: 
\[
\int_{-\infty }^{\infty }dx\,e^{i\xi x}\int dy\,e^{-i\xi \left( x-y\right)
/2}g\left( x-y\right) =-\int_{-\infty }^{\infty }dx\,%
{\displaystyle{\partial  \over \partial x}}%
\left( 
{\displaystyle{e^{i\xi x} \over i\xi }}%
\right) \int dx\,e^{-i\xi \left( x-y\right) /2}g\left( x-y\right) 
\]
\begin{equation}
=-\left. \left( 
{\displaystyle{e^{i\xi x} \over i\xi }}%
\right) \int dx\,e^{-i\xi \left( x-y\right) /2}g\left( x-y\right) \right|
_{x=-\infty }^{x=\infty }+\int_{-\infty }^{\infty }dx\,%
{\displaystyle{e^{i\xi x} \over i\xi }}%
e^{-i\xi \left( x-y\right) /2}g\left( x-y\right) \text{.}
\end{equation}
The first term is zero for our concrete case. The integral may be calculated
explicitly \cite{Gra}. Because the current is conserved, we can take any
particular value of $y$. Choosing $y=0$ the above integral becomes: $%
\int_{-\infty }^{\infty }dx\,%
{\displaystyle{e^{i\xi x/2}g\left( x\right)  \over i\xi }}%
$. Applying this scheme to \ref{fa}, the first correction to the current
density is: 
\[
\left\langle j^{\left( 1\right) }\right\rangle _{\mu }=-%
{\displaystyle{ie \over 2\hbar }}%
\left( \mu _{L}-\mu _{R}\right) \left\{ 
{\textstyle{2 \over i\left( 2\pi \right) ^{3/2}}}%
\int_{-\infty }^{\infty }d\xi \,%
{\displaystyle{\tilde{V}\left( \xi \right)  \over i\xi ^{2}}}%
\int_{-\infty }^{\infty }dx\,%
{\displaystyle{e^{i\xi x/2}\left[ \sin \left( k_{L}x\right) +\sin \left( k_{R}x\right) \right]  \over 2\pi x}}%
\mathop{\rm Si}%
\left( 
{\displaystyle{\xi x \over 2}}%
\right) \right. 
\]
\[
+%
{\textstyle{1 \over 2\left( 2\pi \right) ^{3/2}}}%
\int_{-\infty }^{\infty }d\xi \,%
{\displaystyle{\tilde{V}\left( \xi \right)  \over i\xi ^{2}}}%
\int_{-\infty }^{\infty }dx\,\left\{ 
\mathop{\rm Ci}%
\left[ \left( k_{L}-\xi /2\right) x\right] -%
\mathop{\rm Ci}%
\left[ \left( k_{L}+\xi /2\right) x\right] \right\} 
{\displaystyle{e^{i\xi x/2} \over i\pi x}}%
\]
\begin{equation}
\left. +%
{\textstyle{1 \over 2\left( 2\pi \right) ^{3/2}}}%
\int_{-\infty }^{\infty }d\xi \,%
{\displaystyle{\tilde{V}\left( \xi \right)  \over i\xi ^{2}}}%
\int_{-\infty }^{\infty }dx\,\left\{ 
\mathop{\rm Ci}%
\left[ \left( k_{R}-\xi /2\right) x\right] -%
\mathop{\rm Ci}%
\left[ \left( k_{R}+\xi /2\right) x\right] \right\} 
{\displaystyle{e^{i\xi x/2} \over i\pi x}}%
\right\} \text{.}  \label{fin}
\end{equation}
We concentrate now at: 
\[
\int_{-\infty }^{\infty }dx\,%
{\displaystyle{e^{i\xi x/2}\sin \left( k_{L}x\right)  \over x}}%
\mathop{\rm Si}%
\left( 
{\displaystyle{\xi  \over 2}}%
x\right) =\int_{-\infty }^{\infty }dx\,%
{\displaystyle{i\sin \left( %
{\displaystyle{\xi  \over 2}}x\right) \sin \left( k_{L}x\right)  \over x}}%
\mathop{\rm Si}%
\left( 
{\displaystyle{\xi x \over 2}}%
\right) 
\]
\[
=i\int_{-\infty }^{\infty }dx\,%
{\displaystyle{\cos \left( \left( k_{L}-\xi /2\right) x\right) -\cos \left( \left( k_{L}+\xi /2\right) x\right)  \over 2x}}%
\mathop{\rm Si}%
\left( 
{\displaystyle{\xi x \over 2}}%
\right) 
\]
\[
=%
{\displaystyle{i \over 2}}%
\int_{-\infty }^{\infty }dx\,%
{\displaystyle{\partial  \over \partial x}}%
\left\{ 
\mathop{\rm Ci}%
\left[ \left( k_{L}-\xi /2\right) x\right] -%
\mathop{\rm Ci}%
\left[ \left( k_{L}+\xi /2\right) x\right] \right\} 
\mathop{\rm Si}%
\left( 
{\displaystyle{\xi x \over 2}}%
\right) 
\]
\[
=%
{\displaystyle{i \over 2}}%
\left. \left\{ 
\mathop{\rm Ci}%
\left[ \left( k_{L}-\xi /2\right) x\right] -%
\mathop{\rm Ci}%
\left[ \left( k_{L}+\xi /2\right) x\right] \right\} 
\mathop{\rm Si}%
\left( 
{\displaystyle{\xi x \over 2}}%
\right) \right| _{-\infty }^{\infty } 
\]
\begin{equation}
-%
{\displaystyle{i \over 2}}%
\int_{-\infty }^{\infty }dx\,\left\{ 
\mathop{\rm Ci}%
\left[ \left( k_{L}-\xi /2\right) x\right] -%
\mathop{\rm Ci}%
\left[ \left( k_{L}+\xi /2\right) x\right] \right\} 
{\displaystyle{\sin \left( %
{\displaystyle{\xi x \over 2}}\right)  \over x}}%
\text{.}
\end{equation}
Finally: 
\begin{equation}
\int_{-\infty }^{\infty }dx\,%
{\displaystyle{e^{i\xi x/2}\sin \left( k_{L}x\right)  \over x}}%
\mathop{\rm Si}%
\left( 
{\displaystyle{\xi x \over 2}}%
\right) =%
{\displaystyle{-1 \over 2}}%
\int_{-\infty }^{\infty }dx\,%
{\displaystyle{e^{i\xi x/2} \over x}}%
\left\{ 
\mathop{\rm Ci}%
\left[ \left( k_{L}-\xi /2\right) x\right] -%
\mathop{\rm Ci}%
\left[ \left( k_{L}+\xi /2\right) x\right] \right\}
\end{equation}
and one can see now that in \ref{fin} terms are canceling one each other,
the final result being zero. During our calculations, we have changed the
order of integration by applying the Fubini's theorem. As is pointed out in 
\cite{Re}, we have to be carefully especially with oscillatory integrands.
With a little effort, one can show that we are in the case where the
Fubini's theorem can be applied.

\section{Appendix D}

In this section we will calculate the formulas used in section IV. By using
the field anti-commutation relations it follows: 
\begin{equation}
\left\{ \Psi \left( \left( \Pi _{L}-\Pi _{R}\right) x\right) \,,\,\Psi
^{\dagger }\left( y\right) \right\} =%
{\displaystyle{1 \over V}}%
\sum_{k_{1}>0}e^{ik\left( x-y\right) }-%
{\displaystyle{1 \over V}}%
\sum_{k_{1}<0}e^{ik\left( x-y\right) }=\left( \Pi _{L}-\Pi _{R}\right)
\left( y,x\right)
\end{equation}
and in the infinite limit: 
\[
\left\{ \Psi \left( \left( \Pi _{L}-\Pi _{R}\right) x\right) \,,\,\Psi
^{\dagger }\left( y\right) \right\} =%
{\displaystyle{1 \over \left( 2\pi \right) ^{3}}}%
\int d^{3}k\,\left( \chi _{>}\left( k_{1}\right) -\chi _{<}\left(
k_{1}\right) \right) \,e^{ik\left( x-y\right) } 
\]
\begin{equation}
=%
{\displaystyle{-1 \over i\pi \left( x_{1}-y_{1}\right) }}%
\delta \left( x_{\bot }-y_{\bot }\right) \text{,}
\end{equation}
where we have denoted by $x_{\bot }$ the coordinates: $x_{2}$ and $x_{3}$.
Also, 
\[
\left\langle \Psi ^{\dagger }\left( x\right) \Psi \left( y\right)
\right\rangle _{\mu }=\sum_{k}\Theta \left( k^{2}/2m-\mu _{L}\,\chi
_{>}\left( k_{1}\right) -\mu _{R}\,\chi _{<}\left( k_{1}\right) \right) 
{\displaystyle{e^{-ik\left( x-y\right) } \over V}}%
\]
\begin{equation}
=\left( \Pi _{L}E_{\left[ 0,\mu _{L}\right] }+\Pi _{R}E_{\left[ 0,\mu _{R}%
\right] }\right) \left( x,y\right) \text{.}
\end{equation}
In the infinite limit, 
\begin{equation}
\left\langle \Psi ^{\dagger }\left( x\right) \Psi \left( y\right)
\right\rangle =\int 
{\displaystyle{d^{3}k \over 2\pi }}%
\,\Theta \left( k^{2}/2m-\mu _{L}\,\chi _{>}\left( k_{1}\right) -\mu
_{R}\,\chi _{<}\left( k_{1}\right) \right) e^{-ik\left( x-y\right) }\text{.}
\end{equation}
In consequence: 
\[
\left\langle \left[ \rho _{a}\left( x\right) ,\rho \left( y\right) \right]
\right\rangle _{\mu }=i%
\mathop{\rm Im}%
{\displaystyle{-\delta \left( x_{\bot }-y_{\bot }\right)  \over i\pi \left( x_{1}-y_{1}\right) }}%
\int 
{\displaystyle{d^{3}k \over 2\pi }}%
\,\Theta \left( k^{2}/2m-\mu _{L}\,\chi _{>}\left( k_{1}\right) -\mu
_{R}\,\chi _{<}\left( k_{1}\right) \right) e^{-ik\left( x-y\right) } 
\]
\[
=%
{\displaystyle{i\delta \left( x_{\bot }-y_{\bot }\right)  \over \pi \left( x_{1}-y_{1}\right) }}%
\int 
{\displaystyle{d^{3}k \over 2\pi }}%
\,\Theta \left( k^{2}/2m-\mu _{L}\,\chi _{>}\left( k_{1}\right) -\mu
_{R}\,\chi _{<}\left( k_{1}\right) \right) \cos k_{1}\left(
x_{1}-y_{1}\right) 
\]
\[
=%
{\displaystyle{i\delta \left( x_{\bot }-y_{\bot }\right)  \over \pi \left( x_{1}-y_{1}\right) }}%
\int_{0}^{k_{L}}%
{\displaystyle{dk \over 2\pi }}%
\,k^{2}\,2\pi \int_{0}^{\pi /2}d\theta \sin \theta \,\cos \left( k\left(
x_{1}-y_{1}\right) \cos \theta \right) 
\]
\[
+%
{\displaystyle{i\delta \left( x_{\bot }-y_{\bot }\right)  \over \pi \left( x_{1}-y_{1}\right) }}%
\int_{0}^{k_{R}}%
{\displaystyle{dk \over 2\pi }}%
\,k^{2}\,2\pi \int_{\pi /2}^{\pi }d\theta \sin \theta \,\cos \left( k\left(
x_{1}-y_{1}\right) \cos \theta \right) 
\]
\[
=%
{\displaystyle{i\delta \left( x_{\bot }-y_{\bot }\right)  \over \pi \left( x_{1}-y_{1}\right) ^{2}}}%
\left\{ 
{\displaystyle{k_{L}\cos k_{L}\left( x_{1}-y_{1}\right) +k_{R}\cos k_{R}\left( x_{1}-y_{1}\right)  \over x_{1}-y_{1}}}%
\right. 
\]
\[
\left. -%
{\displaystyle{\sin k_{L}\left( x_{1}-y_{1}\right) +\sin k_{R}\left( x_{1}-y_{1}\right)  \over \left( x_{1}-y_{1}\right) ^{2}}}%
\right\} \text{.} 
\]
We conclude: 
\begin{equation}
f\left( x,y\right) =%
{\displaystyle{i\delta \left( x_{\bot }-y_{\bot }\right)  \over \pi }}%
\left\{ k_{L}^{4}\,u\left( k_{L}\left( x_{1}-y_{1}\right) \right)
+k_{R}^{4}\,u\left( k_{R}\left( x_{1}-y_{1}\right) \right) \right\} \text{,}
\end{equation}
where $u\left( x\right) =%
{\displaystyle{\cos x \over x^{3}}}%
-%
{\displaystyle{\sin x \over x^{4}}}%
$.

\end{document}